\documentclass[twocolumn,showpacs,preprintnumbers,amsmath,amssymb]{revtex4}

\usepackage{graphicx}% Include figure files
\usepackage{dcolumn}% Align table columns on decimal point
\usepackage{bm}% bold math

\begin{document}
\title{Topological gap states of  semiconducting  armchair graphene ribbons }
\author{Y. H. Jeong, S. C. Kim and S. -R. Eric Yang\footnote{corresponding author, eyang812@gmail.com}  }
\affiliation{  Department of Physics, Korea  University, Seoul,
Korea\\}

\begin{abstract}
In semiconducting armchair graphene ribbons  a tensile strain can induce  pairs of topological gap states with
opposite energies. Near the critical value of the deformation
potential these kink and antikink states become almost degenerate with zero
energy and have a fractional charge of one-half. Such a
semiconducting armchair ribbon represents a one-dimensional
topological insulator with nearly zero energy end  states. Using data
collapse of numerical results we find that the shape of the kink
displays an anomalous power-law dependence on the width of the local
lattice deformation. We suggest that these gap states may be probed
in optical measurements. However, "metallic" armchair graphene
ribbons with a gap induced by many-electron interactions have no gap
states and are not topological insulators.
\end{abstract}
\maketitle

%\section{Introduction}

\begin{figure}[!hbpt]
\begin{center}
\includegraphics[width=0.35\textwidth]{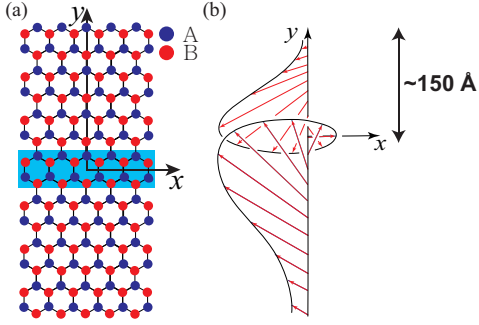}
\caption{(a) Armchair ribbon with length $L_y$ and transverse width
$L_x$. $A$ and $B$ denote carbon atoms.  In the stripe region tensile
strain that has reflective symmetry about the $x$ axis is applied along the ribbon direction. (b) In the stripe
region the pseudospin density $\vec{\Sigma}(y)$ of a gap state
rotates about the $y$ axis with only $x$ and $z$ components. It
represents a kink. Outside the stripe region the length of the
pseudospin goes slowly to zero.} \label{stripe}
\end{center}
\end{figure}

{\it Introduction}.-  Recently a rapid progress has been made in the
fabrication of  armchair graphene ribbons\cite{Novo,Zha,Cai,Kato,Kat}.
They have a great potential for spintronic applications, where
electronic many-body  interactions may play a significant
role\cite{Yang,Lee}. Armchair ribbons may also provide numerous
interesting issues in fundamental physics, such as topological
objects.  By perturbing the graphene lattice and the tunneling
coupling constants  one can give rise to effective gauge and scalar
fields\cite{Kane} that generate solitonic
objects\cite{Jac,Hou,Sas,Oli,Hee}. Generation mechanism of gap and
vortices is at the center of the study of graphene
material\cite{Kat,Pac0,Jac,Cha,Hou,Pac1}.

According to tight-binding calculations an armchair ribbon has a gap
and is  semiconducting  when the transverse width is $L_x=(3M+1)a_0$
or $3Ma_0$, where $a_0$ is the unit cell length and $M$ is an
integer.   But when  $L_x=(3M+2)a_0$ the armchair ribbon has no gap
and is metallic. However, the inclusion of many-electron
interactions produces a gap\cite{Yang}.  It is unclear whether these
armchair ribbons are topological insulators.

In polyacetylene a fermion mass potential produces a gap, and a
twist in this scalar potential produces a zero energy soliton  with
mixed chirality and fermion fractionalization\cite{Hee,Jac,Hou,Oli}.
Similarly,  a chiral gauge field can simulate the effects of
armchair edges and can open a gap. However, the Berry phase is
absent in armchair ribbons (this can be shown by integrating  the
tightbinding wavefunction over the Brillouin zone). It is thus
unclear whether zero energy solitons can exist in such a case, i.e.,
whether the armchair ribbon represents a one-dimensional topological
insulator\cite{topoins}. The purpose of the present paper is to find
a distortion of the chiral gauge field that will produce a kink,
antikink, soliton, and charge fractionalization, as the twist of a
scalar potential does in polyacetylene.

Topological properties can be displayed in the variation of the
pseudospin.  Graphene armchair ribbons consist of A-carbon and
B-carbon atoms, see Fig.\ref{stripe}. The sublattice spin is defined
so that when  an A-carbon atom is occupied its value is 'up' and when
a B-carbon atom is occupied it is 'down' (massless fermions can either
have their sublattice spin pointing along their direction of
wave vector or opposed to it). The valley spin can be defined
similarly. In an armchair ribbon  the sublattice and valley degree
of freedom mix, and their spins can be put together using $4\times
4$ {\it pseudospin} matrices\cite{com1}
\begin{eqnarray}
\vec{\Sigma}=I\otimes\vec{\sigma},
\end{eqnarray}
where  $\vec{\sigma}=(\sigma_{x},\sigma_{y},\sigma_{z})$ are Pauli
spin matrices. We define the pseudospin  density $
\Sigma_i(y)=\Psi^{*}(\vec{r}) \Sigma_i \Psi(\vec{r})$, where
$\Psi(\vec{r})$ is an eigenstate of the Hamiltonian.  A quantum
state with finite occupation probabilities only on A-carbon ('up
spin') atoms to the left of the stripe and only on B-carbon ('down
spin') atoms to the right would constitute a topological object
called a kink.

\begin{figure}[!hbpt]
\begin{center}
\includegraphics[width=0.35\textwidth]{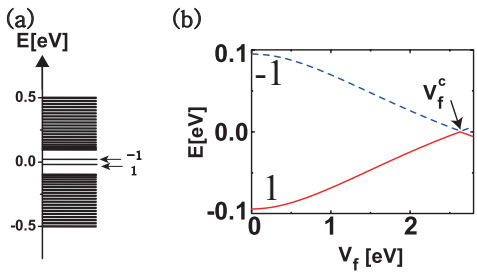}
\caption{(a) Eigenenergies for stripe width $w=5${\AA}, the number
of $k$-space grid points $N=41$, and the strength of deformation
potential $V_f=2$eV (in this paper $L_x=25a_0$ and $L_{y}=1504${\AA}
unless stated otherwise.  Here $a_0=2.46${\AA} is the length of the
unit cell). Label $1 (-1)$ in the gap denotes kink (antikink) with
lower (higher) energy. (b) Kink and antikink energies as a function
of deformation potential $V_f$ for $L_x=25a_0$.  Near the critical
value $V_f^c$ they become almost degenerate. For larger (smaller) values of
$L_x$ the energy  gap is smaller (larger).} \label{energydia}
\end{center}
\end{figure}

We find that the tensile strain with reflective symmetry about the $x$ axis can
generate a pair of gap states with opposite energies when it is
applied along the longitudinal direction of the ribbon  in the stripe
region $-w/2<y<w/2$, see Fig.\ref{stripe}. We find that the properties of these
states cannot be described entirely by any one soliton model (there
are several models, namely, the well-known free Klein-Gordon
equation with a potential with a degenerate vacuum\cite{Jac}, the
polyacetylene model\cite{Hee}, and Jackiw-Rebbi solutions at the
interface between regions of positive and negative
masses\cite{Jac}).

{\it Deformation induced chiral vector potential}.- In the presence
of a deformation induced chiral gauge vector potential $\vec{A}_f$
the effective mass Hamiltonian of two-dimensional Dirac electrons
takes the form
\begin{eqnarray}
H_{\mathbf{K}}=v_{F}\vec{\sigma}\cdot
(\vec{p}-\frac{e}{c}\vec{A}_f(\vec{r}))
\end{eqnarray}
for the $\mathbf{K}$ valley, and
\begin{eqnarray}
H_{\mathbf{K'}}=v_{F}\vec{\sigma}'\cdot
(\vec{p}+\frac{e}{c}\vec{A}_f(\vec{r}))
\end{eqnarray}
for the $\mathbf{K'}$ valley ($e>0$). Here we have  chosen the valley
vectors as $\mathbf{K}=(\frac{4\pi}{3a_0},0)$ and
$\mathbf{K'}=(-\frac{4\pi}{3a_0},0)$.  The $x$ and $y$ components of
$\vec{\sigma}'$ are $-\sigma_{x}$ and $\sigma_{y}$. Note that,
unlike for the real vector potential, the chiral vector potential
appears with opposite signs\cite{chiral} for $\mathbf{K}$ and
$\mathbf{K'}$ valleys  (it is chiral in the sense that it
distinguishes valleys). Since $\mathbf{K}$ and $\mathbf{K'}$ valleys
are coupled by the armchair edges these two $2\times2$ Hamiltonians
are put together as a $4\times4$ Hamiltonian, and the resulting
Hamiltonian is denoted as $H_0$\cite{Brey,Lee}.   For a semiconducting
ribbon the lowest energy conduction subband has the quantized value
$k_x=\frac{\pi}{3L_x}$\cite{Lee} (the metallic ribbon has $k_x=0$
and will be treated separately later) The gap $2\Delta$ separating
conduction and valence bands can be computed from
$\Delta=\frac{\sqrt{3}}{2}t a_0 k_x=0.098$eV (the hopping parameter
is $t= 2.7$eV).

The tensile strain will induce
changes of C-C bonds along the ribbon direction, in addition to
changes of the C-C bonds along other directions. In the effective mass approximation\cite{Sas,tight} it produces
the chiral vector potential $\vec{A}_f(\vec{r})=(A_f\Theta(y),0,0)$,
where $\Theta(y)=1$ in the stripe region  and $0$ outside it, and $\frac{ev_F}{c}A_{f,x}=\delta\gamma_1-\frac{1}{2}(\delta\gamma_2+\delta\gamma_3)$
($\delta\gamma_1$  represents the change of the hoping integral  of the bonds along the ribbon direction
while $\delta \gamma_2$ and $ \delta \gamma_3$ represent those of side bonds).
There is no z-component
of $\vec{A}_f(\vec{r})$, and the y-component is also zero:
$\frac{ev_F}{c}A_{f,y}=\frac{\sqrt{3}}{2}(\delta \gamma_2-\delta \gamma_3)=0$ since $\delta \gamma_2=\delta \gamma_3$ for our tensile strain.
The z-component of the induced magnetic field is non-zero  only at the
edges of the stripe $y=\pm w/2$: $ B_f=A_f(
\delta(y+w/2)-\delta(y-w/2))$. This chiral vector potential leads to
the  correction\cite{ corr}
\begin{eqnarray}
V=-V_f\Theta(y) \Sigma_x \label{chiralvec}
\end{eqnarray}
in the Hamiltonian. The strength of the deformation is
$V_f=\frac{eA_f v_{F}}{c}>0$ and has units of energy.  Increasing
$V_f$ corresponds to weakening of the C-C bonds.

An eigenstate of the total Hamiltonian $H_0+V$ can be solved as a
linear combination of the basis states $\psi_{s,k_y}(\vec{r})$ that
are solutions of the Dirac Hamiltonian $H_0$ of  the armchair
ribbon: $\Psi(\vec{r})=\sum_{s,k_y} C_{s,k_y}\psi_{s,k_y}(\vec{r})$,
where $C_{s,k_y}$ are the expansion coefficients of the lowest
energy conduction  subband ($s=1$) and highest energy valence
subband ($s=-1$). The momentum space  interval $-k_c\leq k_y\leq
k_c$ is divided into $N$ grid points (periodic boundary condition is
used with $k_c=\frac{\pi}{L_y}N$, where $N$ is an odd number and $L_y$ is the length of the ribbon). The
results presented below  are obtained by this method, or by a
tight-binding calculation including many-electron interactions.

\begin{figure}[!hbpt]
\begin{center}
\includegraphics[width=0.5\textwidth]{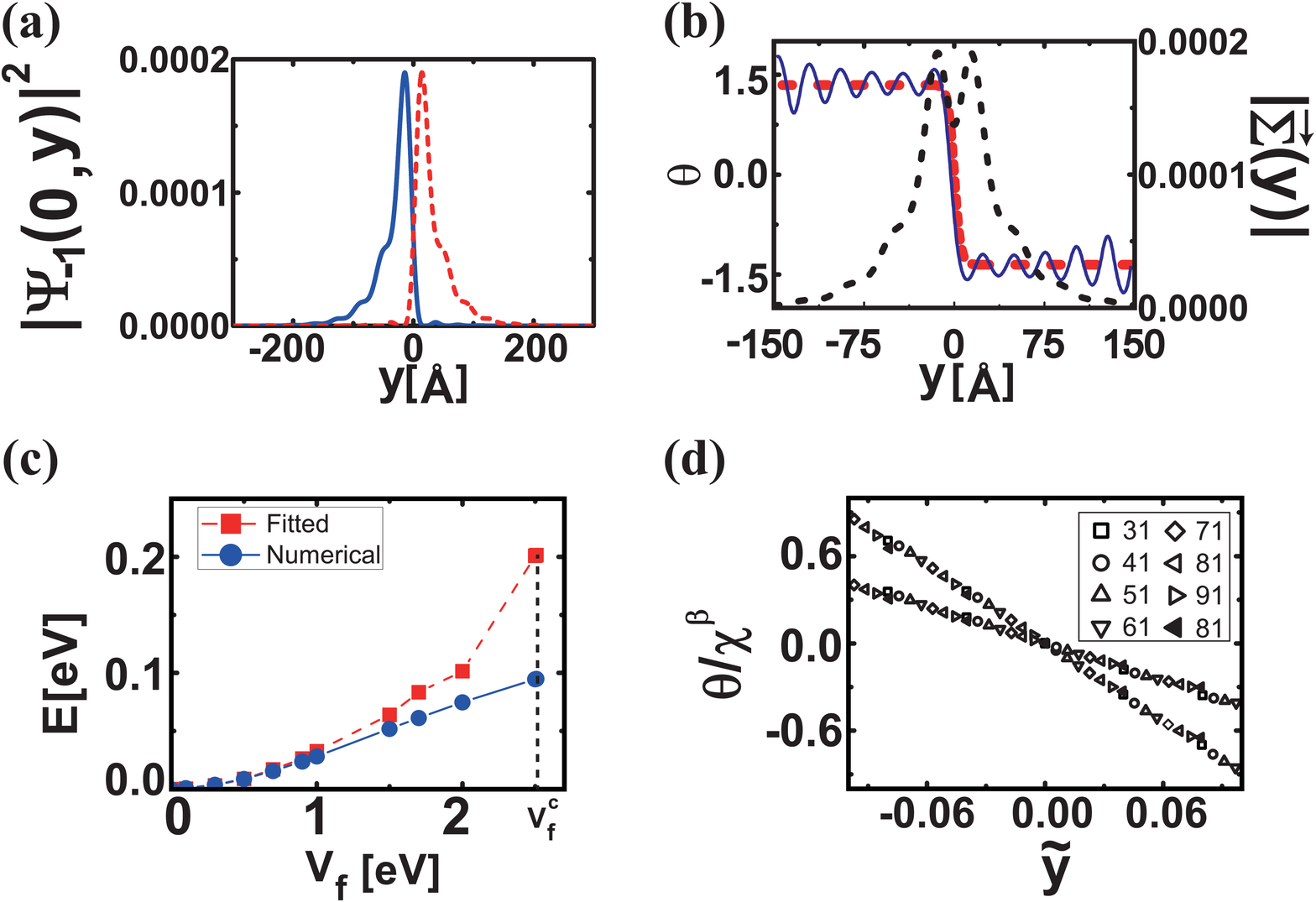}
\caption{(a) Probability density of an   antikink state with energy
$E=0.020$eV. The A component (solid line) and B-component(dashed line)
are shown for $N=81$ and $V_f=2$eV. (b) Numerical result of
$\theta(y)$ (solid line) for the antikink state of (a). Red dashed
line is the fitted result using the amplitude $v=-1.35$ and inverse
length $\alpha=0.143${\AA}$^{-1}$. Also displayed is the magnitude
of $|\vec{\Sigma}(y)|$ (dotted line).  (c) Comparison between
numerical and analytical energies is displayed. (d) Data collapse of
numerical
 data.  Results are for  the stripe widths $w=5${\AA} (open symbols)
and $w=7${\AA} (filled symbols). Strengths of deformation potential
are $V_f=0.87$eV (smaller slope) and $2$eV (steeper slope). Numbers
 in the box indicate numbers of $k$-space grid points $N$ ($k_c$ is
 proportional to $N$). } \label{BvsE3}
\end{center}
\end{figure}

{\it Kink, antikink, and soliton of semiconducting armchair
ribbons}.- We show that kink and antikink states can exist in the
gap. We solve the effective mass Hamiltonian $H_0$ in the presence
of the deformation potential and find that the gap states $ \Psi_1$
and $\Psi_{-1}$ are present with opposite energies, see in
Fig.\ref{energydia}. The wave-function components of $\Psi_1$ and
$\Psi_{-1}$ are related through chiral operation:
$\Psi_1=(\Psi_A,\Psi_B,\Psi_A',\Psi_B')^T$ , and
$\Psi_{-1}=(\Psi_A,-\Psi_B,\Psi_A',-\Psi_B')^T$, where the $A$- and
$B$-components satisfy $\Psi_A(y)=-\Psi_B(-y)$ and the overall phase
factors are removed. The probability density of an antikink state
$\Psi_{-1}$ is shown in Fig.\ref{BvsE3}(a). We have computed    the
angle $\theta$ between the $x$ axis and $\vec{\Sigma}(y)$, see
Fig.\ref{BvsE3}(b) (The $y$ component of the pseudospin density is
$\Sigma_y(y)=0$).
%The winding number
%$w=\frac{1}{2\pi}\int_{-L_y/2}^{L_y/2} \theta(y) dy $ is a non-zero
%integer when integrated over the entire length $L_y$.
The magnitude
of the pseudospin density $|\vec{\Sigma}(y)|$ decays exponentially,
see Fig.\ref{BvsE3}(b). In this exponentially decaying region the
angle is well approximated by $ \theta(y)=\pm v \tanh(\alpha y)$.
 Here the minus sign
corresponds to the antikink state while the plus sign corresponds to the kink state.
In the strong coupling region $V_f>2$ eV the amplitude is
$v\approx \pi/2$.

The energy of a kink is determined by the values of $v$ and
$\alpha$, which  are related to the physical properties of the kink,
namely its amplitude and the inverse size.  For $V_f/V_f^c< 2$ our
numerical energies agree approximately with the analytical result
\begin{eqnarray}
E_{-1}=\frac{\Delta}{2}-\frac{8}{3}(v\alpha)\xi v\,\mathrm{and} \,
E_{1}=-\frac{\Delta}{2} +\frac{8}{3}(v\alpha)\xi v
.\label{kinkenergy}
\end{eqnarray}
In Fig.\ref{BvsE3} (c) the energy correction
$\frac{8}{3}(v\alpha)\xi v$\cite{Jac} is shown and compared with the
numerical result. The parameter $\xi=0.150$eV{\AA} can be determined
by fitting the analytical energy $E$ to the numerical energy.   The
deviation between the numerical and the analytical energies is the
largest near the critical  value $V_f^c=2.51$eV, which is the
maximum value of strength of the deformation potential $V_f^c\approx
t$. Using a tight-binding approach we include  the effects not
included in our model, and  find  more than two gap states.

%\begin{center}
%\includegraphics[width=0.5\textwidth]{chiral.eps}
%\caption{ Degenerate zero-energy chiral modes plotted at $x=0$ as a
%function of $y$.  A- and B-types are shown (a) and  (b). Parameters
%are $N=81$ and $V_f=V_f^c$. } \label{degemode} \label{degenerate}
%\end{center}
%\end{figure}

Near the maximum  value of strength of the deformation potential
$V_f^c$  two nearly  {\it zero} energy
kink and antikink gap states $\Psi_1$ and $\Psi_{-1}$ exist, see
Fig.\ref{energydia}(b). We will call these kink and antikink {\it
solitons}. Their probability densities are similar to the one shown
in Fig.\ref{BvsE3} (a). These modes are nonchiral.
The usual charge counting argument gives
that the sum of the charges of the two gap states is one, implying that
each has a fractional charge of  one half\cite{Jackiw}. In fact, they
are similar\cite{cut} to the nearly degenerate  end states of a one-dimensional topological
insulator with a {\it finite} length\cite{Kit}, despite that the Berry phase is zero in the unperturbed
semiconducting armchair ribbon.

%\begin{figure}[!hbpt]
%\begin{center}
%\includegraphics[width=0.3\textwidth]{scaledphase.eps}
%\caption{ Data collapse of numerical data.  Results are for  the
%stripe widths $w=5${\AA} (open symbols) and $w=7${\AA} (filled
%symbols). Strengths of deformation potential are $V_f=0.87$eV
%(smaller slope) and $2$eV (steeper slope). Numbers in the box
%indicate numbers of $k$-space grid points $N$ ($k_c$ is proportional
%to $N$).} \label{scaling}
%\end{center}
%\end{figure}

{\it Anomalous scaling}.-  We find that the field $\theta(y) $
satisfies near $y=0$ the following finite-size scaling\cite{Park}
relation as a function of the width $w$ and the momentum cutoff
$k_c$
\begin{eqnarray}
\frac{\theta(y)}{\chi^{\beta}}=-A \tilde{y} ,\label{scalingeq}
\end{eqnarray}
where $\tilde{y}=\frac{yk_c}{2\pi}$ is the dimensionless coordinate
and  $\chi=\frac{w k_c }{2\pi}$ is the dimensionless width.  From
our data collapse, shown in  Fig.\ref{BvsE3}(d), and the scaling
relation Eq.(\ref{scalingeq}) we find that the slope of $\theta(y)$
displays an anomalous dependence on the width $w$:
\begin{eqnarray}
v\alpha=A(V_f)\frac{k_c}{2\pi} (\frac{wk_c}{2\pi})^{\beta}
\label{expon}
\end{eqnarray}
with the numerical value of $\beta=0.203$. The anomalous
exponent\cite{Gold} $\beta$ may also be measured experimentally near
$V_f=2$eV, where the amplitude is $v\approx \pi/2$.   The value of
the slope $v\alpha$ can be determined from the measured values of
the kink and antikink energies, given by Eq.(\ref{kinkenergy}). Then
a plot of the slope $v\alpha$ as a function of the width $w$ would
determine the exponent.  Also we find that the constant $A$
increases as the strength of the deformation potential increases.
However, as the transverse length $L_x$ increases, $A$ is
approximately unchanged.

%\begin{figure}[!hbpt]
%\begin{center}
%\includegraphics[width=0.5\textwidth]{tightbinding.eps}
%\caption{Tightbinding DOS of   armchair ribbons near $V_f=V_f^c$,
%where the C-C bonds are cut and two zigzag edges are generated. (a)
%"Metallic" armchair ribbon with $L_x=4.98\AA$ and $L_y=254.2$\AA in
%the presence of many-electron interactions (on-site interaction is
%$U=2.6t$). They remove the zero-energy states. (b) Semiconducting
%armchair ribbon with $L_x=24.9$\AA, $L_y=215.8$\AA, and
%$U=2.6t$.}\label{dos}
%\end{center}
%\end{figure}

{\it Metallic armchair ribbon}.- The inclusion of many-electron
interactions produces a gap in a metallic armchair
ribbon\cite{Yang}. To assess qualitatively the interplay between the
chiral vector potential and  many-electron interactions we model it
in  a half-filled  Hubbard model of a honeycomb lattice\cite{Pis}.
%Many-body correlations are included by choosing the groundstate
%trial wavefunction as  a linear combination of unrestricted
%Hartee-Fock states.
Applying the unrestricted Hartee-Fock method  near $V_f=V_f^c$,
where the C-C bonds are cut, we find a significant gap when the
on-site interaction $U$ is sufficiently large but no gap states
(localized states). This is because the on-site interaction $U$
shifts zero-energy states into the energy continuum where
wavefunctions become delocalized. The Kekule-like distortion,
described by $\gamma^0\Delta$\cite{Hou}, also produces a gap in the
metallic armchair ribbon.  However, adding a chiral vector potential
to it does not generate gap states. The metallic armchair ribbon is
thus not a topological insulator. In contrast, gap states exist for
the semiconducting armchair ribbon when $U$ is not too large.
%For comparison, the DOS of a semiconducting ribbon is also
%shown in Fig.\ref{dos}(b), which has several zero-energy gap states.

{\it Discussions and summary}.-  The optical transition from a kink
$\Psi_1$ to antikink $\Psi_{-1}$ has a substantial optical strength
$|\langle \Psi_1|j_y|\Psi_{-1}\rangle|^2$ when photons are polarized
along the armchair ribbon ($j_y$ is the current operator). It
decreases as $V_f$ increases. However, when photons are polarized
along  the transverse direction  of the ribbon it is zero since
$|\langle \Psi_1|j_x|\Psi_{-1}\rangle|^2=0$. The Coulomb self-energy
and vertex corrections\cite{Kim} renormalize  the bare transition
energy upwards about $15\%$.  Optical resonant oscillations between
$\Psi_1$ and $\Psi_{-1}$ would be also interesting to observe near
$V_f^c$.

We have investigated gap states of  semiconducting armchair ribbons
within an effective mass Hamiltonian model, supplementing it with a
tight-binding calculation including many-electron interactions. A
chiral deformation can form  kink and antikink pairs with opposite
energies. Although the Dirac singularity at zero energy is destroyed
by the gap the zero energy solitons, with a fractional charge of
one half, can exist at the critical  value of the deformation
potential.  Semiconducting armchair ribbons can thus represent a
one-dimensional topological insulator.   Using data collapse of
numerical results we find that the shape of the kink displays an
anomalous power-law dependence on the width of the local lattice
deformation. In contrast   "metallic" armchair ribbons with a
many-body induced gap do not have gap states and are not topological
insulators. We propose optical absorption measurements to explore
the properties of a kink, antikink, and solition of semiconducting
armchair ribbons.   Scanning tunneling microscopy might also be used
to identify the local density of states of the topological gap
states. Fractionally charged fermions can also exhibit non-Abelian braiding statistics\cite{Klinovaja}.

\begin{acknowledgements}
This research was supported by Basic Science Research Program
through the National Research Foundation of Korea (NRF) funded by
the Ministry of Science, ICT and Future Planning (MSIP)
(NRF-2012R1A1A2001554). In addition this research was supported by a
Korea University grant.
\end{acknowledgements}


\begin{references}
\bibitem{com1}Its components satisfy the SU(2) algebra
$ [\Sigma_i,\Sigma_j]=2i\epsilon_{ijk}\Sigma_k$. Note also that they satisfy
$\Sigma_i^2=1$  for $i=x,y,z$. The square of the expectation value of the
pseudospin is $\langle \vec{\Sigma}^2 \rangle=3$ since $\Sigma_i^2=1$.
\bibitem{chiral} It can be written as $\gamma_5\vec{A}_f(r)$ in the chiral representation for the Dirac matrices.
\bibitem{tight} For ribbons with short widths tight-binding approach is more accurate than the Dirac equations.  Even
in this case the tensile strain will generate topological properties of the gap states.
\bibitem{corr}  There is also a scalar potential  term of the
form $-V_f\Theta(y)e^{-2iK_x x} \tau_x\otimes \sigma_x$, which
produces intervalley mixing. ($\tau_x$ is a Pauli matrix for valley spins).  For the derivation of this term see Ref.\cite{Sas} and references therein.
 This term contains the rapidly oscillating factor $e^{-2iK_x x}$ and can be neglected, which  can be shown easily
by calculating its matrix elements using the eigenstates of the armchair ribbon.
\bibitem{cut} Our ribbon has a finite length $L_y$.  We  use the periodic boundary condition along the ribbon direction so a transverse cut produces one finite length armchair ribbon with two zigzag edges.   On the other hand, if we cut an infinitely long ribbon two  semi-infinite ribbons will be generated.
Our ribbon displays  the properties of finite length one-dimensional topological insulators, see   Ref.\cite{Kit}.
\end{references}
\end{document}